\documentstyle[12pt]{article}
\def\ang{{^\circ}}

\def\kpc{\,{\rm kpc}}

\def\cmm2{{\,\rm cm^{-2}}}
\def\cm2{{\,{\rm cm}^2}}
\def\cmm3{{\,{\rm cm}^{-3}}}
\def\gcmm3{{\,{\rm g\,cm^{-3}}}}
\def\kms{\,{\rm km\,s^{-1}}}
\def\Msol{{M_\odot}}

\def\la{\mathrel{\mathpalette\fun <}}

\def\fun#1#2{\lower3.6pt\vbox{\baselineskip0pt\lineskip.9pt
  \ialign{$\mathsurround=0pt#1\hfil##\hfil$\crcr#2\crcr\sim\crcr}}}
\begin{document}
\pagestyle{empty}
\begin{center}
\bigskip


\vspace{.2in}
{\Large \bf Gravitational Microlensing and the Galactic Halo}
\bigskip

\vspace{.2in}
Evalyn I.~Gates,$^{1,2}$ Geza Gyuk,$^{2,3}$ and Michael S. Turner$^{1,2,3}$\\
\vspace{.2in}
{\it $^1$Department of Astronomy \& Astrophysics\\
Enrico Fermi Institute, The University of Chicago, Chicago, IL~~60637-1433}\\

\vspace{0.1in}
{\it $^2$NASA/Fermilab Astrophysics Center\\
Fermi National Accelerator Laboratory, Batavia, IL~~60510-0500}\\

\vspace{0.1in}
{\it $^3$Department of Physics\\
The University of Chicago, Chicago, IL~~60637-1433}\\

\end{center}

\vspace{.3in}

\centerline{\bf ABSTRACT}
\bigskip

By means of extensive galactic modeling we study the
implications of the more than eighty microlensing events that
have now been observed for the composition of the dark halo of
the Galaxy, as well as for other properties of the Galaxy.  We take
the Galaxy to be comprised of luminous and dark disk components,
a bulge, and a dark halo consisting of both MACHOs and cold dark
matter with each component being described by several
observationally motivated parameters.  We pare down an initial
model space of millions of galactic models to viable models,
those which are consistent with the observational data,
including rotation curve, local projected mass density, and
microlensing rates toward the LMC and bulge.  On the basis of a
conservative, minimal set of observational constraints an
all-MACHO halo cannot yet be excluded, although in most viable
models of the Galaxy the halo MACHO fraction is between 0\% and
30\%, consistent with expectations for a universe whose primary
component is cold dark matter.  An all-MACHO halo is required to be
light, and when data on the local escape velocity and
satellite-galaxy proper motions, which probe the extent of the
dark halo, are taken into account, models which have a high
MACHO mass fraction are ruled out.  We also explore the
possibility that there are no MACHOs in the halo.  Finally, we
point out several important tests that could definitively
exclude an all-MACHO e.g., optical depth for microlensing which
is less than $1.5\times 10^{-7}$ toward the LMC or greater than
$3\times 10^{-6}$ toward the bulge.

\newpage
\pagestyle{plain}
\setcounter{page}{1}
\section{\bf Introduction}

Paczynski's bold proposal \cite{pac} to use microlensing to
probe the Galactic halo for dark compact baryonic objects
(referred to as MACHOs for Massive Astrophysical Compact Halo
Objects) has become a reality.  Three collaborations, EROS, OGLE
and MACHO have reported over eighty microlensing events towards
the Galactic Bulge and eight in the direction of the Large
Magellanic Cloud (LMC) \cite{eros,ogle,MACHO,MACHObulge}.  (Preliminary
analyses of the second year MACHO data toward the LMC indicate
two new events \cite{lmcrumor}.)  This detection of microlensing
has opened up a new window for exploring the dark halo of our
galaxy.  In this paper we use the existing data to shed light on
the composition of the Galactic halo.  Some of our key
results have been summarized elsewhere \cite{prl,apjlett}; here
we present the full details of our analysis.

There is compelling evidence that spiral galaxies are imbedded
in extended non-luminous halos. This includes flat rotation
curves measured for almost 1000 spiral galaxies, studies of
binary galaxies including our own galaxy and M31, weak
gravitational lensing, flaring of neutral hydrogen in the disks
and studies of disk warping \cite{haloevid,bbs}.  While the halo of
our own galaxy is in many respects more difficult to study,
there is much important data here too; e.g., the rotation curve
has been measured between 4$\kpc$ and 18$\kpc$, the flaring of
hydrogen gas has been studied, and the orbital motions of
globular clusters and satellite galaxies have been determined
\cite{ourhalo}. All of these support the hypothesis of an
extended dark halo.

Although there is strong evidence for the existence of a
Galactic halo, there is little direct information concerning its
composition.  Since the halos of spiral galaxies are large and
show little sign of having undergone dissipation they can be
expected to reflect the composition of the Universe as a whole,
though perhaps with some biasing (severe in the case of hot dark
matter), and thus their composition is of more
universal importance.  X-ray observations rule out a hot,
gaseous halo, and the Hubble Space Telescope has placed tight
limits on the contribution of faint stars \cite{hst}.  The most
promising candidates for the halo material are baryons in the
form of MACHOs and cold dark matter (CDM) particles.

A baryonic halo invokes the fewest hypotheses: Brown dwarves are
known to exist. Further, substantial baryonic dark matter must
exist given the robust nucleosynthesis lower bound on
$\Omega_B$\cite{turnerschramm}.  However, the success of CDM
models in explaining the formation of large-scale structure and
the appeal of a flat universe and the nucleosynthesis bound to
$\Omega_B$ make a strong case for CDM.  If the bulk of the
Universe exists in the form of CDM, it is inevitable that our
halo contains a significant CDM component \cite{gates}.  (Even
in the most radical scenario for the formation of the Galaxy,
infall onto a baryonic seed mass, the amount of CDM accreted is
at least equal to the total baryonic mass of the galaxy.)
Conclusively demonstrating that the halo is not composed solely
of baryons would comprise additional strong support, albeit
circumstantial, for a halo comprised of CDM particles.

Gravitational microlensing provides a valuable tool for probing
the baryonic contribution to the halo---and of the structure of
the Galaxy itself.  We shall focus on measurements of the
optical depth for microlensing (the probability that a given
distant star is being microlensed).  The optical depth is
determined by the amount and distribution of mass in microlenses
along the line of sight.  With sufficient lines of sight a sort
of galactic tomography could in principle be performed.  At
present only a few lines of sight have been probed: several in
the direction of the LMC, which probe the halo, and several in
the direction of the Bulge, which probe the inner galaxy.  The
small probability for microlensing, of order $10^{-6}$, means
that millions of stars must be monitored.  There are many fields
of view available in the direction of the Bulge and so
tomography of the inner galaxy is a realistic possibility.  The
situation for probing the halo is not as promising: with available resources
only the direction toward the LMC has star
fields of sufficiently high density to be useful. However, a
space-based search should be able to target the Small Magellanic
Cloud (SMC), perhaps some of the larger globular clusters and
the closer galaxies such as M31.

Even with precise knowledge of the optical depths toward the LMC
and bulge, it would still be difficult to interpret the results
because of the large uncertainties in the structure of the
Galaxy. As it is, small number statistics for the LMC lead to a
range of optical depths further complicating the analysis.
Detailed modeling of the Galaxy is essential to drawing reliable
conclusions.

Thus, we adopt the following strategy for determining the MACHO
composition of our galactic halo.  We construct models of the
Galaxy with five components: luminous and dark disks, baryonic
and CDM halos, and a bulge.  We describe each by parameters
whose values are allowed to vary over a range motivated by
previous modeling and observations. By simultaneously varying
all the parameters we construct a very large space of models
(more than ten million); from this we find a subspace of viable
models consistent with the diverse set of observations that
constrain the Galaxy---rotation curve, local projected mass
density, measurements of the the amount of luminous matter in
the disk and bulge, and measurements of the optical depth for
microlensing toward the bulge and the LMC.  The distribution of
the MACHO halo fraction in these viable models allows us to
infer its preferred value. Further, since it is difficult to
exclude an all-MACHO halo we focus attention on models where the
MACHO fraction is high to see what observations might be crucial
in testing this possibility.

Our approach is not the only one that could be pursued. The
MACHO Collaboration has focused on a handful of representative
galactic models that are meant to span the larger range of
possibilities \cite{newMACHO}.  This allows them to study each
model in more detail and address not only the number of
microlensing events, but also their durations (which are
determined by a combination of the MACHO mass, distance and
velocity across the sky). They reach a similar conclusion
concerning the MACHO fraction of the halo---it is small in most
models of the Galaxy---though they construct a model with an
all-MACHO halo.  While their approach allows them to address the
question of the masses of MACHOs, they do not constrain their
models with the totality of observations and thus they cannot
address the viability of the models they consider.  Indeed, we
find their all-MACHO model incompatible with the observational
data.

A few caveats should be kept in mind. Because the acceptance of
the MACHO and EROS experiments to event duration are limited,
the present data address only the halo component made up of
MACHOS with masses from about $10^{-7}M_\odot$ to $10^2
M_\odot$.  It has been argued that objects of mass outside this
range are unlikely: MACHOs of mass $10^{-7} M_\odot$ evaporate
on a time scale less than the age of the galaxy \cite{derujula};
Black holes of greater than $10^4 \Msol$ would disrupt the
globular clusters \cite{carr}. However, there remains the
possibility that the halo baryons are in the form of either
molecular clouds with a fractal distribution \cite{depaolis} or
very massive ($m \sim 10^2 M_\odot - 10^4 M_\odot$) black holes
\cite{carr}. Neither of these options is particularly
compelling---molecular clouds should have collapsed by the
present and the massive progenitors of such black holes would
likely have produced $^4$He or heavy elements---however, they
cannot be ruled out conclusively at this time.

In our analysis we also assume that MACHOs are smoothly
distributed rather than clumped. If they were strongly clumped
the microlensing rate could vary significantly across the sky,
which might appear to allow a smaller or larger optical depth
toward the LMC for a given MACHO halo fraction.  However, if
more than one clump were on average expected in a patch of sky
the size of the LMC then the optical depth would be again close
to its average. Thus, for clumping to significantly affect the
optical depth there must be at most a few clumps in the solid
angle subtended by the LMC.  But if this is the case, then we
can expect no more than a few thousand such clumps over the
entire sky out to the distance of the LMC. To be a significant
fraction of the total halo mass ($\sim {\rm few} \times 10^{11}
M_{\odot}$) each clump must be of order ${\rm few} \times 10^{8}
M_{\odot}$, far greater than the mass of a globular cluster. A
few thousand of these objects residing in the halo would seem to
be ruled out firmly by dynamical constraints based upon the
stability of the disk \cite{carr}.

Our paper is organized as follows: In the next Section we
discuss galactic modeling and the minimal constraints we impose
on models.  In Section 3 we discuss the implications of
microlensing on galactic modeling.  We also consider additional
reasonable constraints, the local escape velocity and satellite
galaxy proper motions, which preclude any model with an
all-MACHO halo.  In Section 4 we examine more closely the few
models that allow an all-MACHO halo (within the minimal
constraints) as well as those models that allow a no-MACHO halo.
In the final Section we summarize our results and discuss future
observations---from measurements of galactic parameters to
strategies for the microlensing measurements---that can sharpen
conclusions concerning the MACHO fraction of the halo.

\section{\bf Galactic Modeling}

Modeling of the Galaxy is an established subject---the basic
features and dimensions of the Galaxy were determined early in
this century---but also one that is still undergoing significant
change.  Evidence for a dark halo has accumulated over the past
two decades (see, e.g.  \cite{glxmodels}) and over the past five
years or so a strong case has for a bar-like, rather than
axisymmetric, bulge has developed \cite{manybars}.  Microlensing
has the potential for contributing significantly to our
understanding of the structure of the Galaxy, both of the
composition of the halo and the mass distribution interior to
the solar circle.

The current picture of the Galaxy is a barred spiral,
consisting of three major components: a central bulge (bar), a
disk and a dark halo. The luminous components are a thin,
double exponential disk with a vertical scale height of about
$0.3\kpc$ and a radial scale length of about $3.5\kpc$, a
smaller (few percent of the disk mass) ``thick'' disk with
vertical scale height of about $1\kpc$ to $1.5\kpc$ \cite{GWK},
and a central bulge region, which recent observations indicate
is a triaxial bar \cite{manybars}.

Evidence for the dark halo is less direct, but firm nonetheless.
It comes from the rotation curve, which is flat out to at least
$18\kpc$ (and probably out to $50\kpc$) and the approach of
Andromeda and the Galaxy toward one another.  At the solar
circle about 40\% of the centripetal acceleration is provided by
the gravitational force of the halo, and beyond that the
fraction is even greater.  The mass of the Galaxy inferred from
the approach of Andromeda is at least a factor of ten greater
than that which can be accounted for by stars alone
\cite{haloevid}.  Moreover, the evidence for dark halos
associated with spiral galaxies in general is very secure.  A
recent survey of the rotation curves of more than 900 spiral
galaxies indicates flat or slightly rising rotation curves at
the limit of the observations, providing strong evidence for
their massive dark halos \cite{persic2}.  {}From a completely
different direction, Brainerd, Blandford and Smail \cite{bbs}
have mapped the dark halos of several spiral galaxies by means
of their weak-gravitational lensing of very distant galaxies.
Their results indicate that the halos studied have radial extent
of at least $100 h^{-1}\kpc$ and total masses in excess of
$10^{12} M_{\odot}$.

The values of the parameters that describe the components of the
Galaxy are not well determined; this is especially true for the
halo whose presence is only known by its gravitational effects.
In addition, there is interplay between the various components
as the observations typically constrain the totality of the
model, rather than a given component.  Modeling uncertainties
introduce significant, irreducible uncertainties in the
determination of the MACHO content of the halo.  In order to
understand these uncertainties we explore a very wide range of
models that are consistent with all the data that constrain the
Galaxy.

We consider two basic models for the bulge, the first following
Dwek et al. \cite{dwek} who have utilized DIRBE surface
brightness observations to construct a triaxial model for the
bulge:
\begin{equation} \rho_{\rm BAR} = {M_0 \over 8 \pi abc}
e^{-s^2/2}, \qquad s^4 = \left[ {x^2\over a^2} + {y^2\over b^2}
\right] ^2 + {z^4 \over c^4} ,
\end{equation}
where the bulge mass $M_{\rm Bulge} = 0.82 M_0$, the scale
lengths $a = 1.49\kpc$, $b = 0.58\kpc$ and $c = 0.40\kpc$, and
the long axis is oriented at an angle of about $10^{\circ}$ with
respect to the line of sight toward the galactic center.  While
we do not take the axes and inclination angles to be modeling
parameters, we later explore the sensitivity of our results to
them.  We also consider an axisymmetric Kent model for the bulge
\cite{kent}. The rotation curve contribution was calculated in
the point mass approximation. At $r=5\kpc$ this approximation is accurate to
better than $10\%$.

The bulge mass is not well determined, and we consider $M_{\rm
Bulge}=(1 - 4) \times 10^{10} M_\odot$, in steps of $0.5\times
10^{10}M_\odot$.  Previous estimates have been in the range $(1
- 2) \times 10^{10} M_\odot$ \cite{glxmodels,kent,zhao},
although a recent study by Blum \cite {blum} which utilized the
tensor virial theorem found a bar mass closer to $3 \times
10^{10} M_\odot$ (assuming a bar orientation of 20 degrees --
smaller (larger) angles of orientation imply larger (smaller)
bulge masses).

For the disk component we take the sum of a ``fixed,'' thin
luminous disk whose constituents (bright stars, gas, dust, etc.)
are not expected to serve as lenses,
\begin{equation}
\rho_{\rm LUM}(r,z) = {\Sigma_{\rm LUM}\over 2h}\,
\exp [-(r-r_0)/r_d] e^{-|z|/h},
\end{equation}
with scale length $r_d = 3.5\kpc$, scale height $h = 0.3\kpc$,
and local projected mass density $\Sigma_{\rm LUM} = 25
M_\odot\,{\rm pc}^{-2}$ \cite{lumdisk}, and a ``variable'' disk
component whose constituents are assumed to be lenses.  For the
variable component we consider first a distribution similar to
that of the luminous matter but with varying scale lengths $r_d
= 3.5\pm 1 \kpc$, and thicknesses $h=0.3\kpc$, and $1.5\kpc$.
We also consider a model where the projected mass density varies
as the inverse of galactocentric distance (Mestel model)
\cite{1/r}.

The motions of stars perpendicular to the galactic plane have been
used to infer the {\it total} local projected mass density
within a distance of $0.3\kpc - 1.1\kpc$ of the galactic plane
\cite{sigma0}.  The values so determined are between
$40M_\odot\,{\rm pc}^{-2}$ and $85M_\odot\,{\rm pc}^{-2}$.  As a
reasonable range we require that $\Sigma_{\rm TOT}(1\kpc ) =
\int_{-1\kpc}^{1\kpc}\rho (r_0, z)dz = 35 - 100 M_\odot\,{\rm
pc}^{-2}$, which constrains the local projected mass density
of the dark disk to be $10M_\odot\,\le \Sigma_{\rm VAR}
\le 75 M_\odot \,{\rm pc}^{-2}$.   (We also include
the contribution of the halo to $\Sigma_{\rm TOT}(1\kpc )$,
which for flattened halo models can be significant, about $20
M_\odot\,{\rm pc}^{-2}$, and reduces the mass density that the
variable disk can contribute.)

The dark halo is assumed to be comprised of two components,
baryonic and non-baryonic, whose distributions are independent.
We first assume independent isothermal distributions for MACHOs
and cold dark matter with core radii $a_i = $2, 4, 6, ..., 18,
20$\kpc$,
\begin{equation}
\rho_{{\rm HALO},i} = {a_i^2+r_0^2 \over a_i^2 +r^2}\, \rho_{0,i}\ ,
\label{eq:iso}
\end{equation}
where $i=$ MACHO, CDM and $\rho_{0,i}$ is the local
mass density of component $i$.

There are indications from both observations \cite{Sackflat,rix}
and CDM simulations \cite{quinn} that halos are significantly
flattened.  In order to explore the effects of flattening we
also consider models with an axis ratio $q = 0.4$ (an E6 halo)
for both the baryonic and non-baryonic halos with distributions
of the form
\begin{equation}
\rho_{{\rm HALO},i} = {a_i^2+R_0^2 \over a_i^2 +R^2 + (z/q)^2}\,
\rho_{0,i}\ ,
\end{equation}
where $(R,z)$ are cylindrical coordinates.  While flattening
does affect the local halo density significantly, increasing it
by roughly a factor of $1/q$ (see Ref.~\cite{apjlett}), it does not
affect the halo MACHO fraction significantly.

Finally, we consider the possibility that the MACHOs are not
actually in the halo, but instead, due to dissipation, are more
centrally concentrated.  To describe this we use the
distribution in Eq.~(\ref{eq:iso}) but with $r^2$ replaced by
$r^n$, for $n=3,4$ and core radii $a_{\rm MACHO} =1,2 \kpc$.
Such a distribution approximates models of a spheroidal
component \cite{glxmodels,giudice} (note, in these models we
also explicitly include a Dwek bar).

We construct our models of the Galaxy by letting the parameters
describing the various components vary independently.  By doing
so we consider millions of models.  We pare down the space of
models to a smaller subset of viable models by requiring that
observational constraints be satisfied.  The kinematic
requirements for our viable models are: circular rotation speed at
the solar circle ($r_0 = 8.0\kpc \pm 1\kpc$) $v_c =220\kms\pm
20\kms$; peak-to-trough variation in $v(r)$ between $4\kpc$ and
$18\kpc$ of less than 14\% (flatness constraint \cite{gates});
and circular rotation velocity at $50\kpc$ greater than
$150\kms$ and less than $307\kms$. We first impose this minimal set
of constraints in order to be as conservative as possible
in our conclusions; later we impose additional reasonable, but
less secure constraints, involving the rotation curve at large
distances and the local escape velocity.

We also impose constraints from microlensing, both toward the
bulge and toward the LMC. The optical depth for microlensing a
distant star by a foreground star is \cite{griest}
\begin{equation}
\tau = {4 \pi G\over c^2} {\int^\infty_0 ds \rho_s (s)
\int^s_0 dx \rho_l (x) {x(s-x)/ s} \over \int^\infty_0 ds \rho_s (s)},
\end{equation}
where $\rho_s$ is the mass density in source stars, $\rho_l$ is
the mass density in lenses, $s$ is the distance to the star
being lensed, and $x$ is the distance to the lens \cite{lenses}.
In calculating the optical depth toward the bulge, we consider
lensing of bulge stars by disk, bulge and halo objects; for the
LMC we consider lensing of LMC stars by halo and disk objects.
Except where we are constructing microlensing maps of the bulge
(see Section 5) we define the direction of the bulge to be toward
Baade's window, $(b,l) = (-4\ang,1\ang)$.

We adopt the following constraints based upon microlensing data: (a)
$\tau_{\rm BULGE} \geq 2.0\times 10^{-6}$ and (b) $0.2\times
10^{-7}\le \tau_{\rm LMC} \le 2\times 10^{-7}$ \cite{larger}.
The bulge constraint is based upon the results of the OGLE
Collaboration \cite{ogle} who find $\tau_{\rm BULGE} = (3.3\pm
1.2)\times 10^{-6}$, as well as the results of the MACHO
Collaboration who find $\tau_{\rm BULGE} = 3.9 ^{+1.8}_{-1.2}\times 10^{-6}$
\cite{MACHObulge}.  To be sure, there are still important
uncertainties, e.g., detection efficiencies and whether or not
the stars being lensed are actually in the bulge; however, we
believe this to be a reasonable bound to the optical depth.  The
optical depth to the LMC is based upon the MACHO Collaboration's
measurement \cite{MACHOprl}, $\tau_{\rm LMC} = 0.80\times 10^{-7}$,
as well as the results of the EROS Collaboration \cite{eros}.
Here too there are uncertainties. In addition to the obvious
small number statistics, the events might not all be
microlensing. As a reasonable first cut we have taken the 95\%
Poisson confidence interval based upon the MACHO results.

Bulge microlensing provides a crucial constraint to galactic
modeling and eliminates many models.  It all but necessitates a
bar of mass at least $2\times 10^{10}M_\odot$, and, as has been
emphasized by others \cite{zhao}, provides additional evidence
that the bulge is bar-like.  Because of the interplay between
the different components of the Galaxy, the bulge microlensing
optical depth indirectly constrains the MACHO fraction of the halo.
On the other hand, LMC microlensing only constrains the MACHO
fraction of the halo.

\section{\bf Implications of Microlensing for Galactic Modeling}

In this Section we discuss the characteristics of the viable
models, focussing particularly on the composition of the halo
(MACHO fraction and local halo mass density), but also paying
attention to the other parameters in our galactic models.  We
display our results in histograms of the number of viable models
as a function of various modeling and derived parameters.
These plots {\it resemble} likelihood functions that are
marginalized with respect to those parameters.  They are in fact
not likelihood distributions; because the most important
uncertainties in modeling the Galaxy are systematic in
character, e.g., the model of the Galaxy itself, the rotation
curve, the shape of the halo, and even the galactocentric
distance and local speed of rotation, we resisted the urge to
carry out a more rigorous statistical analysis which might have
conveyed a false level of statistical significance.

We first discuss the features of the models that satisfy our
minimal constraints and then go on to discuss the models that
survive when we impose additional constraints that better serve
to define the extent of the dark halo (escape velocity and
rotation curve at large distances as defined by satellite galaxy
proper motions).  In these discussions we rely heavily upon
histograms which detail the characteristics of the acceptable
galactic models. However, before we do, let us summarize our
main results:

\begin{itemize}

\item  In most viable models the halo MACHO fraction is between
0\% and 30\%, though when only the minimal constraints are
applied there are models with MACHO fraction greater than
60\%.  When the additional constraints
are applied there are no viable models with
halo MACHO fraction greater than 60\% (see Fig.~1).
(Halo MACHO fraction $f_B$ is defined to be the MACHO mass
fraction of the halo interior to $50\kpc$).

\item In viable models the local MACHO mass density is
sharply peaked around $10^{-25}\gcmm3$ (see Fig.~2) and the total
MACHO mass (within $50\kpc$) is peaked around $1\times 10^{11}
M_\odot$.

\item In viable models with a flattened halo
the total local halo mass density
is between about $4\times 10^{-25}\gcmm3$ and
$1.5\times 10^{-24}\gcmm3$ (see Fig.~2).  Flattening
increases the local halo mass density by factor of
order the axis ratio.

\item The bulge microlensing constraint precludes any model with
a Kent (axisymmetric) bulge, and the bar mass in most viable
models is between $2\times 10^{10}M_\odot$ and $3\times
10^{10}M_\odot$.  The necessity of a relatively heavy galactic
bar plays an important role constraining the halo MACHO fraction
to a small fraction.

\end{itemize}

\subsection{Minimal constraints}

There are several features that are generic to most models that
satisfy the minimal set of constraints (see Figs.~3-8).  The
most important of these is that independent of almost all the
model parameters, the peak of the MACHO fraction occurs for $f_B
\la 20\%$ (the only exception being a spherical halo model with
a very small core radius for the non-baryonic component, which
peaks at $f_B \sim 30\%$).  While the range of MACHO fraction
extends from 0\% to 90\%, most models have $f_B < 30\%$.  (We
discuss the handful of high MACHO-fraction models in the next
Section).  No model with a thick dark disk (either exponential
or $1/r$ profile) and $f_B > 60\%$ survives our constraints, and
the distribution for these thick disk models peaks at $f_B \sim
0$.  The absence of MACHOs in the halo is allowed because a
thick disk can contribute up to $0.5\times 10^{-7}$ to the
optical depth toward the LMC \cite{prl}, which allows the LMC
microlensing constraint to be satisfied without recourse to
MACHOs in the halo.

The bulge mass in most models is between $2\times 10^{10} M_\odot$
and $3\times 10^{10}M_\odot$, which is consistent with estimates
from recent efforts to model the bar \cite{zhao,blum}.   Models with a Kent
bulge do not provide sufficient microlensing toward the bulge, and as pointed
out in previous work by the authors and others \cite{prl,goulddisk}, the disk
cannot provide more than about $1 \times 10^{-6}$ to the optical depth toward
the bulge. A heavy
bar is necessary to obtain optical depths to the galactic bulge
in excess of $3 \times 10^{-6}$, as currently suggested by the
experimental data. The distribution of galactocentric distance
($r_0$) is somewhat dependent on the disk model, with thick disk
models generally favoring smaller $r_0$.  The distribution for
the local circular velocity is relatively broad, but it is
generally peaked at the low end of the range, around $210\kms
-220 \kms$. The trend for all dark disk models is toward larger
scale length (r$_d$).  The value of the disk surface density
depends on the disk model, although lighter disks are favored in
all cases (i.e., little mass in the dark disk).

The distribution of optical depths toward the LMC and the bulge
are shown in Figs.~3-8.  In general, $\tau_{\rm LMC}$ is
relatively flat. This is easily understood: for a given model,
the microlensing optical depth is sensitive only to the MACHO
fraction, which is unaffected by the kinematic cuts.  For
thick-disk models (both exponential and $1/r$) there is also a
relatively large bin at the smallest allowed value of $\tau_{\rm
LMC}$.  This is due to additional allowed models with very small
halo MACHO fraction where the LMC lensing is done by the disk
(lensing toward the LMC is negligible in thin-disk models
\cite{prl}).  The bulge optical depth is somewhat peaked toward
the low end of the acceptable range, mainly due to the
difficulty of achieving $\tau_{\rm bulge} > 3 \times 10^{-6}$.

The local MACHO mass density peaks at about $10^{-25}
\gcmm3$ in all models and the mass of MACHOs in the
halo peaks at about $1\times 10^{11}M_\odot$.  However, the total local halo
mass density is more dependent on the halo model, in particular on
whether or not the halo is flattened; see Fig.~2.  (Since the MACHO fraction
of the halo is small, this also applies to the local mass
density of CDM particles.)  Flattening of the halo, for
which there is good evidence, increases the local halo
density by a factor of order the axis ratio $q$.  In a flattened halo model,
the local halo density is larger by a factor
\begin{equation}
{\rho_0^{flattened} \over \rho_0^{spherical}} = {\sqrt{1-q^2}\over q \sin^{-1}
(\sqrt{1-q^2}) },
\end{equation}
relative to a spherical halo model with the same asymptotic rotation velocity
and core radius (for the E6 halo,
this factor is about 2.)
This has important implications for the
direct detection of non-baryonic dark matter, and is discussed
in detail elsewhere \cite{apjlett}.  However, our results for the MACHO
fraction of the halo are essentially independent of the amount of halo
flattening as can be seen in figures 3-11.  Both the total mass of the halo and
the MACHO halo mass shift slightly toward smaller values in a flattened halo
model.

\subsection{Additional constraints}

The models we have considered viable thus far have been subject
to a very minimal set of constraints -- that is, we have
tried to be as generous as possible in admitting models, probably
too generous.  There are additional constraints which bear on the
size and extent of the dark halo.  They are especially crucial
to the issue of the MACHO fraction of the halo:  Microlensing
toward the LMC closely constrains the mass of the MACHOs in the
halo, and therefore the halo MACHO fraction depends sensitively
upon the total halo mass.  The models with high MACHO fraction
are characterized by light halos; the additional constraints
place a stringent lower bound to the halo mass and thus upper
bound to the MACHO fraction, eliminating all models with MACHO
fraction greater than 60\%.

The first additional constraint on the galactic potential that
we consider comes from the local escape velocity.  Based upon
the velocity of the fast moving stars Leonard and Tremaine
\cite{leonard} have determined that the local escape velocity
lies in the range $450 \kms < v_{\rm ESC} < 650\kms$ (with 90\%
confidence level), with a stronger lower limit of $430\kms$.
Kochanek \cite{kochanek} obtains a slightly higher range of $489
\kms < v_{\rm ESC} < 730\kms$. Based on these values we adopt
$v_{\rm ESC} > 450\kms$.\footnote{The escape velocity from an
isothermal halo increases logarithmically; to compute $v_{\rm
ESC}$ we truncate the halo at a distance of $100\kpc$.}

Next we consider the information about the galactic rotation
curve at large distances ($50\kpc - 100\kpc$) based upon the
proper motions of satellites of the Milky Way.  Recently Jones,
Klemola and Lin \cite{Jones} have measured the proper motion of
the LMC.  They find a total galactocentric transverse velocity
of $215 \pm 48\kms$.  Proper motions for Pal 3 \cite{Cudworth}
(galactocentric distance $79\kpc$) and Sculptor
\cite{Schweitzer} (galactocentric distance $95\kpc$) have also
been measured, yielding $252\kms \pm 85\kms$ and $199\kms \pm
58\kms$ respectively.  Assuming that these satellite galaxies
are bound to our Galaxy, they provide strong evidence that the
galactic halo is massive and extended.

Finally, a study of the rotation curves of over 900 spiral
galaxies \cite{persic2} indicates that for all of these galaxies
the rotation curves are flat, rising or only gently falling at
twice the optical radius ($r_{\rm opt}\equiv 3.2r_d$), depending
on the luminosity.  Based on rotation curves of galaxies similar
to the Galaxy ($L/L_*=1.4h^{2}, r_d \approx 3.5\kpc$), the
rotation velocity at $2r_{\rm opt} \sim 22\kpc$ should be within
a few percent of $v_c$, and further, at a galactocentric
distance of $50\kpc$ the rotation velocity should be at least
$200\kms$.  Combining this with the satellite proper motions we
require $180\kms \leq v_c(50\kpc) \leq 280\kms$.

We impose these additional constraints on our ``canonical''
model---E6 halo, thin, double-exponential disk, and Dwek
bar---with all other parameters allowed to vary as before.  The
results are displayed in Fig.~9.  The most striking consequence
of the additional kinematic constraints is the exclusion of all
models with a baryon fraction greater than 60\%, and essentially
all models with a baryon fraction greater than 50\%.  It is
worth noting that this result follows from either constraint
alone.  That is, models with an all-MACHO halo are characterized
by {\em both} $v_{\rm ESC}< 450\kms$ {\em and} $v_c(50\kpc)<
180\kms$.  The results for a spherical halo are similar.

The halo MACHO fraction for these models is strongly peaked
around 10\% to 20\%. This result is independent of the bar mass,
local disk surface mass density, disk scale length and our
galactocentric distance.  It is also insensitive to the optical
depth for microlensing toward the galactic bulge.  It is, as one
would expect, sensitive to the optical depth for microlensing
toward the LMC.

These additional constraints also narrow the estimate for the
total mass of the Galaxy (within $50\kpc$) to $(5 \pm 1) \times
10^{11} M_{\odot}$.  This is consistent with the value obtained
recently by Kochanek \cite{kochanek}, who used similar
constraints on the extent of the dark halo, although a much more
restricted set of galactic models.

\section {\bf Very MACHO and No-MACHO Halos}
\subsection{Very-MACHO halos}

In Figs.~3 to 8 the characteristics of galactic models with
MACHO fraction $f_B \ge 0.75$ are shown as dotted lines.  (It
should be noted that the histograms for these models with
very-MACHO halos have been multiplied by a factor of 50 relative
to the other models.)  The crucial common feature of very-MACHO
models is a light halo (total mass less than $4\times
10^{11}M_\odot$).  Only thin-disk models allow $f_B\ge 0.75$.
The reason for this illustrates how the bulge microlensing
constraint also influences other aspects of the galactic model.
Models with an exponential thick disk require a heavier bar to
account for microlensing toward the bulge: A thick disk
contributes far less to microlensing toward the bulge than does
a thin disk \cite{prl}.  On the other hand, the rotation curve
from our position outward requires a heavy disk for support if
the halo is light.  Therein lies the rub: the inner part of the
rotation curve cannot tolerate both a heavy disk and a heavy
bar.

Because very-MACHO models are characterized by light halos they
are also characterized by: (i) a small local rotation speed,
$v_c \le 215\kms$; (ii) large (total) local surface mass
density, $\Sigma_{0} \geq 60 M_\odot\,{\rm pc}^{-2}$; (iii)
light bar, $M_{Bulge} =2.0 \times 10^{10} M_\odot$ in most of
these models; (iv) a rotation curve that falls to a small
asymptotic value, $v(50 \kpc) \la 180\kms$; and (v) a local
escape velocity that is less than $420\kms$.  Further, because
the bar is the most efficient source of lensing, a lighter bar
results in a low optical depth toward the bulge, $\tau_{\rm
bulge}\simeq 2 \times 10^{-6}$.  Finally, to avoid having a halo
that is too light, these models are necessarily characterized by
high optical depth toward the LMC, $\tau_{\rm LMC}\sim 2 \times
10^{-7}$.

\subsection{No-MACHO halos}

Because the optical depth for microlensing toward the LMC is so
much smaller than it would be for an all-MACHO halo one should
also consider the possibility that there are no MACHOs in the
halo.  Further, the optical depth toward the LMC is based on
only three events seen by the MACHO Collaboration and two by the
EROS Collaboration.  Not only are the numbers small, so that
Poisson fluctuations alone are large, but it is not impossible
that some of the events are not even due to microlensing.  In
that regard, the MACHO Collaboration refers to their events as
two candidates and one microlensing event (the amplitude 7
event) \cite{newMACHO}, while the EROS Collaboration has established
that one of their events involves a binary star (of period much
shorter than the event duration) \cite{erosbin}.  Thus, the actual
optical depth could be quite small.

If the optical depth for microlensing toward the LMC is much
less than $10^{-7}$ (the current central value), it could be
explained by a combination of microlensing of LMC stars by LMC
stars \cite{lmclens} and a thick disk component (a thick disk
can contribute up to $0.5\times 10^{-7}$, though it should
be noted that a thick disk cannot also account for the large
microlensing rate toward the bulge).

Another possibility is that the MACHOs responsible for microlensing
toward the LMC are in a more centrally condensed component, e.g.,
the spheroid.  In Figs.~10 and 11 we show the characteristics
of models with a no-MACHO halo and MACHO spheroid with density
profiles $r^{-n}$ ($n=3,4$) and core radii $b=1,2\kpc$.  The
viable models are characterized by:  (i) very small MACHO fraction,
spheroid mass/halo mass less than 0.2; (ii) very low optical
depth, $\tau_{\rm LMC} \la 5\times 10^{-8}$; and (iii) spheroid
mass which peaks at $5\times 10^{10}M_\odot$ for $n=3$ and $3\times
10^{10}M_\odot$ for $n=4$, consistent with
independent dynamical measurements \cite{giudice}.

\section{Discussion and Summary}

\subsection{\bf Microlensing and the bulge}

The number of microlensing events detected in the direction of
the galactic bulge is currently more than eighty and will continue to
grow.  As the statistics improve, the optical depth along different
lines of sight toward the bulge can be determined,
allowing tomography of the inner galaxy, in turn providing
information about the shape, orientation, and mass of the bulge,
and indirectly about the Galaxy as a whole.

Already the unexpectedly high optical depth towards the
galactic center provides further evidence that the bulge is more
bar-like than axisymmetric.  Much more can be learned.  In
Fig.~12 we present microlensing maps of the bulge for several
different models. The first panel shows contours of constant
$\tau_{\rm bulge}$ for a massive ($M_{Bulge} = 4.0\times
10^{10}M_\odot$) Kent bulge with a light disk.  Even with this
very high bulge mass, the microlensing rates are not high enough
to account for the observations.  The second panel shows a
microlensing map for a slightly less massive ($M_{Bulge} = 3.0\times
10^{10}M_\odot$) Dwek bar oriented almost directly towards us,
$\theta = 10\ang$.  Despite the lower mass which makes this
model more likely to pass kinematic cuts, the optical depths are
much higher, with bulge-bulge events clearly dominating.  A
slight asymmetry in galactic longitude is apparent, but it may
be too small to be detected. The third panel shows a microlensing
map for the same mass bar, but oriented at $45\ang$.  The
optical depths for microlensing are much smaller, the contours
are considerably less steep and more elongated along the
longitude axis.  For comparison, the the effect of a heavier
disk is shown in panels four and five, for a models similar to
those in panels two and three.  The additional microlensing
provided by the disk results in higher optical depths and an
elongation of the microlensing contours along the direction of
galactic longitude.

While the orientation of the bar provides a strong signature in
the microlensing maps, the overall rate is an important
constraint by itself.  The models shown in panels three and five
with an orientation of $45^\circ$ are already excluded by our
constraint, $\tau_{\rm bulge}\geq 2.0\times 10^{-6}$.  Figure 13a
shows the number of viable models with a thin disk and flattened
halo as a function of bar orientation.  Clearly a bar pointing
towards us is preferred, with bar orientations of greater than
$30\ang$ almost entirely excluded.

The modeling we have described here has already indicated the
necessity of a relatively massive bar, $(2-3)\times 10^{10}
M_\odot$, even in the case of a bar oriented at $10\ang$ from
our line of sight.  This, together with the results shown in
Fig.~13 suggest that the bar has a mass of $(2-3)\times
10^{10}M_\odot$ and is oriented at an angle of less than $20 -
30\ang$ from our line of sight.  As discussed earlier,
considering rate alone there is a degeneracy between bar mass
and orientation: lower mass can be traded for smaller angle.  As
can be seen in Fig.~12 mapping can break this degeneracy.

\subsection{\bf Future directions}

While the results of the microlensing experiments to date seem
to strongly indicate that the primary component of the halo is
not MACHOs, as we have emphasized here it is not yet possible to
exclude this hypothesis with any certainty.  Since the question
is of such importance, it is worth considering future measurements
that could led to more definite conclusions.  Based upon our
extensive modeling we can identify a a number of key measurements.

Recall that the models with all-MACHO halos had a number of
distinctive features: (i) large optical depth toward the LMC,
$\tau_{\rm LMC} \simeq 2\times 10^{-7}$; (ii) small optical
depth toward the bulge, $\tau_{\rm bulge}\simeq 2 \times
10^{-6}$; (iii) a small local rotation speed, $v_c \le 215\kms$;
(iv) large (total) local surface mass density, $\Sigma_{0} \geq
60 M_\odot\,{\rm pc}^{-2}$; (v) light bar, $M_B \simeq 2.0
\times 10^{10} M_\odot$; (vi) a rotation curve that falls to a
small asymptotic value, $v(50 \kpc) \la 180\kms$; and (vii) a
local escape velocity that is less than $420\kms$.

What then are the prospects for falsifying the all-MACHO halo
hypothesis?  Because $\tau_{\rm LMC}$ is apparently so small, it
may be difficult to accumulate sufficient statistics over the
next few years to exclude the possibility that $\tau_{\rm LMC}$
is as large as $2\times 10^{-7}$.  It may be more promising to
establish that $\tau_{\rm bulge}$ is greater than $2\times
10^{-6}$, due to the higher microlensing rate toward the bulge,
Or, other observations could establish that the mass of the
bulge is in excess of $2\times 10^{10}M_\odot$, which cannot be
tolerated in models where the halo is entirely comprised of
MACHOs.

Several characteristics of an all-MACHO halo involve parameters
of the galactic model and the galactic rotation curve.
Improvements here could be equally decisive.  The study of the
proper motions of satellite galaxies will further constrain the rotation curve
at large distances, and the recent observation of a
dwarf galaxy at a galactocentric distance of $16\kpc$
\cite{sagit} presents yet another opportunity.  Continued
efforts to deduce the local escape velocity might well rule out
all-MACHO scenarios. A more precise determination of the local
circular velocity and position would also help limit the range
of viable models.  Precision measurements of the pulse arrival
times for the binary pulsar PSR 1913+16 are reaching the level
of precision where the effects of solar acceleration, which
depends upon both $r_0$ and $v_c$, can be accurately determined
\cite{taylortiming}.

Equally interesting is testing the hypothesis of a no-MACHO
halo.  Measurements of the event duration and light-curve
distortions due to parallax effects could help discriminate
between MACHOs in the halo and disk and/or LMC.  Likewise, the
distribution of events in the LMC provides an important test of
whether or not the lenses are part of the LMC.  It is probably
more difficult to determine whether or not the lenses are in the
spheroid (as opposed to the halo).

\subsection{\bf Summary}

Microlensing has already proven its utility as a probe of
the structure of the Galaxy.  Based upon the existing data---which
is likely to represent but a small fraction of what will be
available over the next few years---and the extensive modeling
discussed here important conclusions can already be drawn.

First and foremost, the MACHO fraction of the galactic halo in
most viable models of the Galaxy is small---between 0\% and 30\%
(see Fig.~1).  The few models with a halo MACHO fraction of
greater than 60\% are characterized by a very light halo.  When
additional reasonable constraints that define the minimal extent
of the halo (such as local escape velocity and proper motions of
satellite galaxies) are taken into account none of these models
remain viable.  The apparent elimination of the promising
baryonic candidate for the dark matter halo of our own galaxy
further enhances the case for cold dark matter and provides
further impetus for the efforts to directly detect cold dark
matter particles (e.g., neutralinos and axions).

Second, it is not impossible that the halo of the Galaxy
contains no MACHOs.  If the optical depth for microlensing
toward the LMC is at the low end of the credible range, say less
than about $0.5\times 10^{-7}$, the microlensing events seen
could be due to microlensing by objects in the disk and/or LMC.
Or, it could be that the lenses are not halo objects, but rather
exist in a more centrally condensed component of the Galaxy
(e.g., the spheroid).

Third, based upon our modeling we conclude that the plausible
range for the local density of dark halo material is between
$6\times 10^{-25}\gcmm3$ and $13 \times 10^{-25}\gcmm3$, most
which is not in the form of MACHOs (see Fig.~2).  This estimate
is about a factor of two higher than previous estimates because
we have taken the flattening of the halo into account
\cite{apjlett}.

Fourth, it is not possible to account for the large microlensing
rate in the direction of the bulge with an axisymmetric bulge; a
bar of mass $(2-3)\times 10^{10}M_\odot$ is required to meet our
minimal constraint $\tau_{\rm bulge} \ge 2\times 10^{-6}$.

Finally, while we are not able to rule out an all-MACHO halo
with certainty, our modeling points to future measurements that
could be decisive.  The very few models with very-MACHO halos
($f_B \ge 0.75$) that survive our minimal set of constraints
have distinctive features that make allow them to be falsified:
$\tau_{\rm LMC} \simeq 2\times 10^{-7}$; $\tau_{\rm bulge}\simeq
2 \times 10^{-6}$; $v_c \le 215\kms$; $\Sigma_{0} \geq 60
M_\odot\,{\rm pc}^{-2}$; $M_B \simeq 2.0 \times 10^{10}
M_\odot$; $v_c(50 \kpc) \la 180\kms$; $v_{\rm esc}\la 420\kms$.

\section*{Acknowledgments}
We thank C. Alcock, K. Cudworth and D. Bennett for helpful conversations.  This
work was
supported in part by the DOE (at Chicago and Fermilab) and the
NASA (at Fermilab through grant NAG 5-2788).

\newpage

\section*{Figure Captions}
\bigskip

\noindent{\bf Figure 1:} The number of viable models as a function of halo
MACHO fraction for our minimal set of constraints (solid line) and additional
constraints based upon the local escape speed and proper motions of
satellite galaxies (dashed line).

\medskip
\noindent{\bf Figure 2:} The local mass density of halo matter,
(a) total and (b) in MACHOs,  for models with a Dwek bar, thin
double-exponential disk and spherical halo (dotted line) or
flattened (E6) halo (solid line).

\medskip
\noindent{\bf Figure 3:} Galactic model with a Dwek bar, thin
double-exponential disk, and spherical halo. (a) Number of viable models as a
function of the halo MACHO fraction, $f_B$,  for various values of the
model input parameters.  For the histograms labeled $\tau_{LMC}$ and
$\tau_{BULGE}$ only it is assumed that these optical depths are
known a precision of 10\%. (b) Histograms of the number of viable models as a
function of various model input parameters and optical depths.
(c) Histograms of the
number of viable models corresponding to selected model output parameters,
from left to right and top to bottom, local mass density of CDM particles,
local mass density of halo MACHOs, asymptotic rotation velocity,
local escape velocity, total mass of MACHOs in the halo (within 50\,kpc),
and total halo mass (within 50\,kpc).  The dotted lines correspond to
the number of models with $f_B \geq 0.75$, scaled upward by a factor of 50.

\medskip
\noindent{\bf Figure 4:} As in Fig.~3 for a galactic model with a Dwek bar,
thick double-exponential disk, and spherical halo.

\medskip
\noindent{\bf Figure 5:} As in Fig.~3 for a galactic model with a Dwek bar,
1/r disk, and spherical halo.

\medskip
\noindent{\bf Figure 6:} As in Fig.~3 for a galactic model with a Dwek bar,
thin double-exponential disk, and flattened (E6) halo.

\medskip
\noindent{\bf Figure 7:} As in Fig.~3 for a galactic model with a Dwek bar,
thick double-exponential disk, and flattened (E6) halo.

\medskip
\noindent{\bf Figure 8:} As in Fig.~3 for a galactic model with a Dwek bar,
1/r disk, and flattened (E6) halo.

\medskip
\noindent{\bf Figure 9:} As in Fig.~3 for a galactic model with a Dwek bar,
thin double-exponential disk, and E6 halo, where in addition we have imposed
the constraints based upon the local escape speed and proper motions of
satellite galaxies.

\medskip
\noindent{\bf Figure 10:} As in Fig.~3 for a galactic model with a Dwek bar,
thin double-exponential disk, $1/r^3$ baryonic spheroid and spherical
CDM halo.

\medskip
\noindent{\bf Figure 11:} As in Fig.~3 for a galactic model with a Dwek bar,
thin double-exponential disk, $1/r^4$ baryonic spheroid and spherical
CDM halo.

\medskip
\noindent{\bf Figure 12:} Contours of constant optical depth for
microlensing in the direction of the bulge for
(a) Kent model with $M_B = 4.0\times 10^{10}\Msol$ and
$\Sigma_{VAR} = 60 \Msol \,{\rm pc}^{-2}$;
and models with $M_B = 3.0\times 10^{10}\Msol$ and
(b) $\theta = 10\ang$, $\Sigma_{VAR} = 10 \Msol \,{\rm pc}^{-2}$;
(c) $\theta = 45\ang$, $\Sigma_{VAR} = 10 \Msol \,{\rm pc}^{-2}$;
(d) $\theta = 10\ang$, $\Sigma_{VAR} = 60 \Msol \,{\rm pc}^{-2}$;
(e) $\theta = 45\ang$, $\Sigma_{VAR} = 60 \Msol \,{\rm pc}^{-2}$.
Contours are $0.5, 1.0, 2.0, 3.0, 4.0, 5.0, 6.0 \times 10^{-6}$
from the outside inwards.

\medskip
\noindent{\bf Figure 13:} (a)The number of viable models as a
function of bar orientation for models with a thin
double-exponential disk and E6 halo. (b) the microlensing
optical depth to the bar as a function of the bar orientation to
our line of sight.


\begin{thebibliography}  {std}
\bibitem{1/r} Such a model (referred to as a Mestel
disk \cite{binney}) produces a flat rotation curve in the
plane of the galaxy; however, in order to
account for a rotation velocity of $220\kms$, a local surface mass
density of $220 \Msol\,{\rm pc}^{-2}$ is required.
P.~Sackett has recently revisited these models,
because a thin Mestel disk ($h\sim 0.3\kpc$) whose local surface
density is about $220 \Msol\,{\rm pc}^{-2}$ can:
(i) marginally account for the bulge rate; (ii) marginally account for
the LMC rate; and (iii) produce a flat rotation curve ($v = 220\kms$)
in the galactic plane without a halo.  However, such a model would not produce
a flat rotation curve outside the galactic plane nor would it explain
the warping of the disk and the flaring of neutral gas; moreover,
it is in severe conflict with kinematic studies that indicate
$\Sigma_{\rm TOT}(1\kpc )$ is at most $100\Msol\,{\rm pc}^{-2}$
\cite{sigma0}.
\bibitem{lenses}  Our expression for $\tau$ when applied to the
bulge implicitly assumes that all the mass density in the
disk and bulge is available for lensing.  However, the bulge lenses
are certainly not bright stars.  Zhao et al. \cite{zhao}
find that correcting for this is a small effect.  To account
for the fact that the lenses are unlikely to be bright stars
we simply do not include the contribution of the fixed luminous
disk to $\tau_{\rm BULGE}$.
\end{thebibliography}
\end{document}